\RequirePackage{fix-cm}
\documentclass[smallextended]{svjour3}       % onecolumn (second format)
\smartqed  % flush right qed marks, e.g. at end of proof
\usepackage{graphicx}
\usepackage{color}
\usepackage[numbers]{natbib}
\usepackage{amssymb, amsmath}
\usepackage[normalem]{ulem} % Added by author to use strike-out lines

\begin{document}

\title{Material properties of a low contraction and resistivity silicon--aluminum composite for cryogenic detectors%\thanks{Grants or other notes
}

\titlerunning{A low contraction and resistivity silicon--aluminum composite}        % if too long for running head

%\author{First Author         \and
%        Second Author %etc.
%}

\authorrunning{Takekoshi et al.} % if too long for running head

%\institute{F. Author \at
%              first address \\
%              Tel.: +123-45-678910\\
%              Fax: +123-45-678910\\
%              \email{fauthor@example.com}           %  \\
%%             \emph{Present address:} of F. Author  %  if needed

\author{Tatsuya Takekoshi$^{1,2}$, Kianhong Lee$^{2}$, Kah~Wuy Chin$^{2}$, Shinsuke Uno$^{2}$, Toyo Naganuma$^{3}$, Shuhei Inoue$^{2}$, Yuka Niwa$^{4}$, Kazuyuki Fujita$^{5}$, Akira Kouchi$^{5}$, Shunichi Nakatsubo$^{6}$, Satoru Mima$^{7,8}$, Tai Oshima$^{9,10}$}

\institute{
\email{ttakekoshi@mail.kitami-it.ac.jp}\\
1. Kitami Institute of Technology, 165 Koen-cho, Kitami, Hokkaido 090-8507, Japan\\
2. Institute of Astronomy, Graduate School of Science, The University of Tokyo, 2-21-1 Osawa, Mitaka, Tokyo 181-0015, Japan \\
3. Graduate School of Informatics and Engineering, The University of Electro-Communications, Cho-fu, Tokyo 182-8585, Japan \\
4. Tokyo Institute of Technology, 2-12-1 Ookayama, Meguro-ku, Tokyo 152-8551, Japan \\
5. Institute of Low Temperature Science, Hokkaido University, Sapporo 060-0819, Japan \\
6. Institute of Space and Astronautical Science, Japan Aerospace Exploration Agency, Sagamihara 252-5210, Japan \\
7. National Institute of Information and Communications Technology, Kobe, Hyogo 651-2492, Japan\\
8. The Institute of Physical and Chemical Research (RIKEN), 2-1 Hirosawa, Wako-Shi, Saitama 351-0198, Japan\\
9. National Astronomical Observatory of Japan, Mitaka, Tokyo 181-8588, Japan \\
10. The Graduate University for Advanced Studies (SOKENDAI), 2-21-1 Osawa, Mitaka, Tokyo 181-0015, Japan \\
}

\date{Received: date / Accepted: date}

\maketitle

\begin{abstract}
We report on the cryogenic properties of a low-contraction silicon--aluminum composite, namely Japan Fine Ceramics SA001, to use as a packaging structure for cryogenic silicon devices.
SA001 is a silicon--aluminum composite material (75\% silicon by volume) and has a low thermal expansion coefficient ($\sim$1/3 that of aluminum).
The superconducting transition temperature of SA001 is measured to be 1.18~K, which is in agreement with that of pure aluminum, and is thus available as a superconducting magnetic shield material.
The residual resistivity of SA001 is 0.065~$\mathrm{\mu \Omega m}$, which is considerably lower than an equivalent silicon--aluminum composite material.
The measured thermal contraction of SA001 immersed in liquid nitrogen is $\frac{L_{293\mathrm{K}}-L_{77\mathrm{K}}}{L_{293\mathrm{K}}}=0.12\%$, which is consistent with the expected rate obtained from the volume-weighted mean of the contractions of silicon and aluminum. 
The machinability of SA001 is also confirmed with a demonstrated fabrication of a conical feedhorn array, with a wall thickness of $100~\mathrm{\mu m}$. 
These properties are suitable for packaging applications for large-format superconducting detector devices.
\end{abstract}

\section{Introduction}
\label{sec:intro}
Large format arrays of superconducting detectors, such as microwave kinetic inductance detectors and transition edge sensors, are being developed to enhance the survey speed of astrophysical telescopes.
These detectors are often built on silicon wafers, which are packaged with aluminum components like feedhorn antenna arrays and wafer support structures \cite{2018JLTP..193..103J,2021ApJ...922...38M, 2020JLTP..200..384L, 2020SPIE11453E..1FD}.
However, the large difference in thermal contraction between silicon and these metals can result in serious misalignment of the optical components, and thermal stress damaging the device when cooled to cryogenic temperatures.
Although Invar is a solution for differential thermal contraction, it is not widely used because of the performance degradation of superconducting detectors due to ferromagnetism.

Recently, cosmological experiments at millimeter and submillimeter wavelengths (e.g., CLASS \cite{2018SPIE10708E..2PA, 2022RScI...93b4503A, 2020JLTP..199..289D}, AdvACT \cite{2018JLTP..193.1041S}) have introduced the silicon--aluminum composite alloy material, Sandvik-Osprey CE7, produced by the spray-forming method \cite{2022RScI...93b4503A}, which has a low thermal expansion coefficient ($\sim$1/3 that of aluminum).
CE7 also has the advantage of being a non-magnetic and superconducting magnetic shield material owing to its superconducting transition temperature of 1.2~K, which is above the operation temperature of the detector stages ($\lesssim 0.3$~K).

Herein, we report the measurement results for an alternative silicon--aluminum composite material, Japan Fine Ceramics SA001, in terms of thermal contraction, superconducting transition temperature, and electric resistivity.
We also demonstrate the machinability of SA001 with a test cut of a conical feedhorn array.

\section{Material}
\label{sec:material}
SA001\footnote{https://www.japan-fc.co.jp/en/products/cate01/cate0104/sisic75vol-sa001.html} is a silicon--aluminum composite material with a 75:25 volume ratio produced by the infiltration method, which shows a structure of silicon grains ($\sim$10--100~$\mathrm{\mu m}$ in size) filled with aluminum.
SA001 is lighter, more rigid, and has a lower thermal expansion coefficient than aluminum and its alloys, and is thus a functional substitute for aluminum or steel alloys in industrial use.
The physical properties of SA001 at room temperature are summarized in Table~\ref{table:1}.
The significant difference between SA001 and CE7 is in the electrical resistivity, which of SA001 is three times smaller than that of CE7. 
The other properties are almost equivalent to those of CE7.

\begin{table}
\begin{minipage}{10cm}
\caption{Physical properties of CE7 and SA001.}
\label{table:1}
\begin{tabular}{lll}
\hline\noalign{\smallskip}
& CE7$^a$ & SA001  \\
\noalign{\smallskip}\hline\hline\noalign{\smallskip}
\textbf{Room} \\
Composition (wt\% of silicon)  & 70 & 72$^b$ \\
Composition (vol\% of silicon) & 73$^c$ & 75 \\
Thermal expansion coefficient at 200~$^\circ$C (K$^{-1}$) & $\sim$9$\times 10^{-6}$ & 9$\times 10^{-6}$ \\
Density     (g~cm$^{-3}$) & 2.43 & 2.4 \\
Young's modules (GPa)  & 129.2 & 120 \\
Possion's ratio & 0.26 & 0.29 \\
Resistivity at 300~K ($\mathrm{\mu \Omega m}$) & 1.15 & 0.4 \\
\noalign{\smallskip}\hline
\multicolumn{2}{l}{\textbf{Cryogenic}}\\
Transition temperature $\mathrm{T_c}$ (K) & 1.19 & 1.18$\pm 0.01$ \\
Residual Resistivity at 1.5~K ($\mathrm{\mu \Omega m}$) & 0.50 & 0.065$\pm0.001$ \\
Thermal conductivity at 1.2~K ($\mathrm{W m^{-1} K^{-1}}$) & 0.06$^d$ & 0.45$^d$ \\
Thermal contraction $\frac{L_{293\mathrm{K}}-L_{77\mathrm{K}}}{L_{293\mathrm{K}}}$ & 0.1\% & $0.120\pm0.013$\% \\
Thermal contraction $\frac{L_{293\mathrm{K}}-L_{4\mathrm{K}}}{L_{293\mathrm{K}}}$ & 0.1\% & $0.12$\%$^e$ \\
\noalign{\smallskip}\hline
\end{tabular}

\footnotetext[1]{Reference \cite{2022RScI...93b4503A}.}
\footnotetext[2]{Calculated from the volume composition.}
\footnotetext[3]{Calculated from the weight composition.}
\footnotetext[4]{Estimated using the Wiedemann–Franz law from residual resistivity at 1.5~K.}
\footnotetext[5]{Volume-weighted mean of thermal contraction at 4~K of silicon and aluminum.}
\end{minipage}
\end{table}

\section{Cryogenic Measurements}
\subsection{Superconducting transition temperature and resistivity}

\begin{figure}
\includegraphics[width=0.7\textwidth]{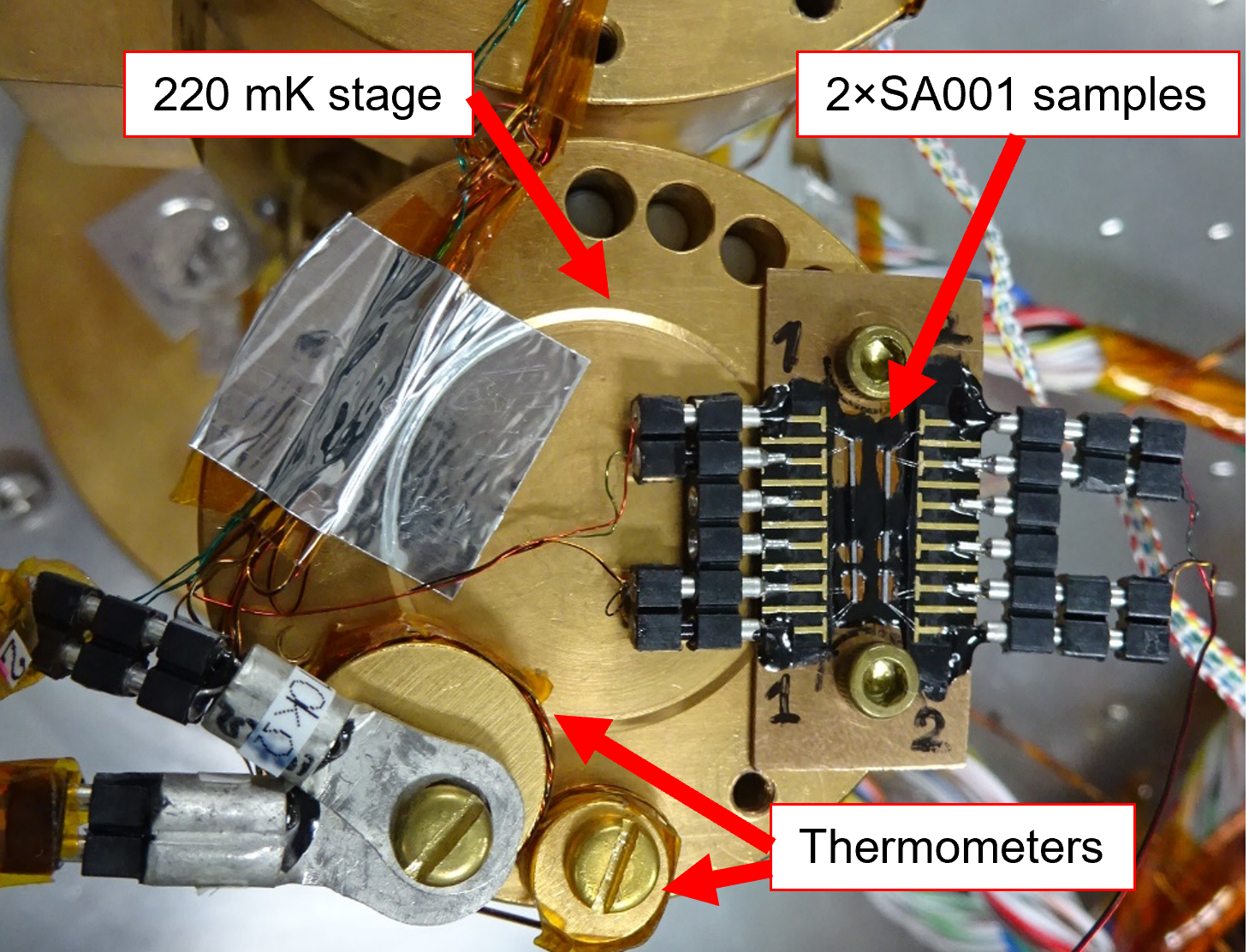}
\caption{Cryogenic measurement configuration to determine transition temperature and resistivity.}
\label{figure:1}
\end{figure}

\begin{figure}
\includegraphics[width=0.5\textwidth]{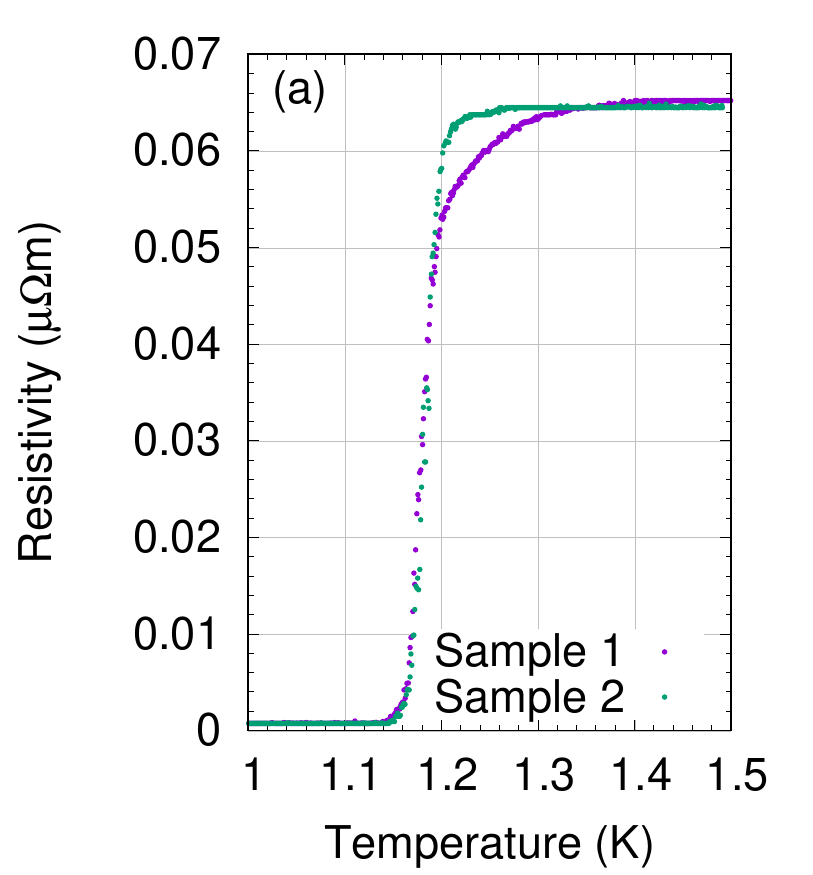}
\includegraphics[width=0.5\textwidth]{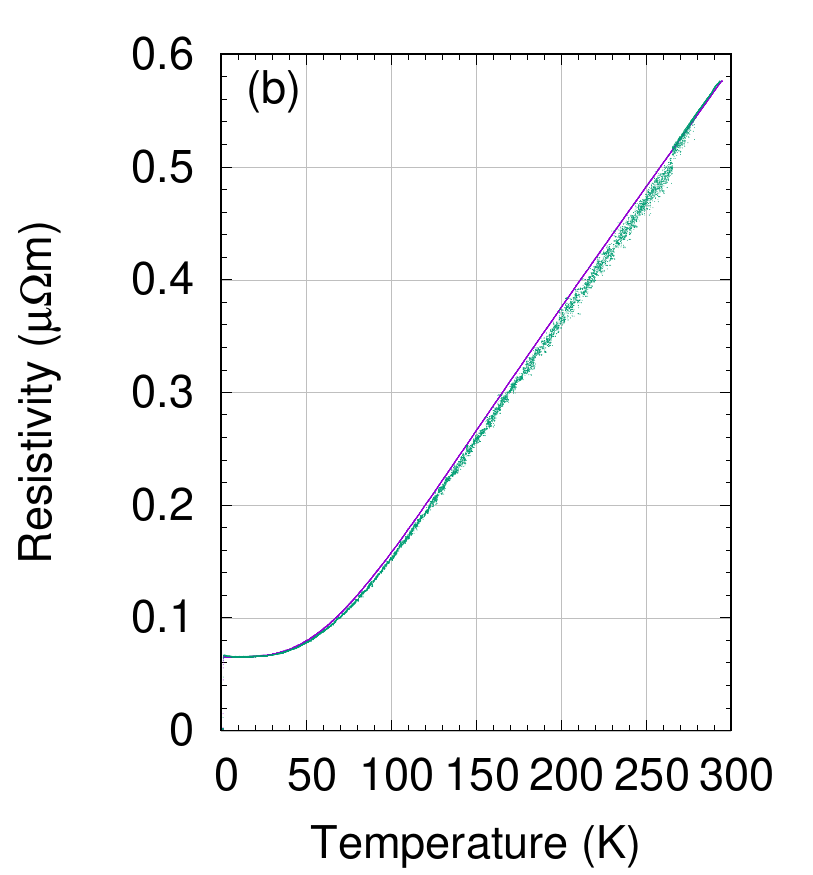}
\caption{Measured resistivity-temperature curves of the SA001 samples (a) around the transition edge temperature and (b) in course of the cooldown from 300~K.}
\label{figure:2}
\end{figure}

We measured the resistivity--temperature curve of SA001 to investigate the superconducting transition temperature and material's electric properties at cryogenic temperature.
Two fine ($0.42\times0.15\times8.5~\mathrm{mm^3}$) samples of SA001 were cooled down to 220~mK on the cold stage of the $^{3}$He sorption refrigerator, Simon Chase He-10 (Figure~\ref{figure:1}).
Resistances of the samples were continuously measured by the 4-terminal method using an AC resistance bridge (AVS-47B) in the course of the cooldown.
The temperature of the samples were measured by well-calibrated thermometers Cernox CX-1010 and Ruthenium oxide RX-102A.
The result is shown in Figure~\ref{figure:2}.
The sharp transitions to the superconducting state at the transition temperature\footnote{$\mathrm{T_c}$ of SA001 is defined at the temperature of half of the normal resistivity, which is consistent with the definition of $\mathrm{T_c}$ of CE7 \cite{2022RScI...93b4503A}.} $\mathrm{T_c}=1.18\pm0.01$~K for both samples assure the reproducibility of the superconducting property of SA001. 
The measurement error is given by the calibration error of the thermometers.
The measured $\mathrm{T_c}$ is equivalent to those of pure bulk aluminum ($\mathrm{T_c}=1.18$~K\cite{ekin2006experimental}) and CE7 ($\mathrm{T_c}=1.19$~K \cite{2022RScI...93b4503A}),
Thus, we conclude that SA001 surroundings can efficiently function as a superconducting magnetic shield at the operating temperature of detectors ($<1$~K).

We also obtained a residual resistivity (RR, defined at 1.5~K) of $0.065\pm0.001~\mathrm{\mu \Omega m}$ for SA001, which is considerably lower than that of CE7 ($0.5~\mathrm{\mu \Omega m}$ \cite{2022RScI...93b4503A}).
Using the Wiedemann--Franz law with the theoretical Lorentz number of $L=2.44\times 10^{-8}~\mathrm{W \Omega K^{-2}}$, we estimate the thermal conductivity at 1.2~K to be $0.45$ and $0.059$~$\mathrm{W m^{-1} K^{-1}}$ for SA001 and CE7, respectively, suggesting that SA001 has eight times better thermal conductivity than CE7. 
Below the transition temperature, thermal conductivity of CE7 at 300~mK has been reported to be 0.005~$\mathrm{W m^{-1} K^{-1}}$ \cite{2022RScI...93b4503A}.
The value is expected to be a lower limit for SA001 below the transition temperature.
Therefore, SA001 can be cooled from room to the operation temperature of detectors very efficiently.

\subsection{Thermal contraction}
\begin{figure}
\includegraphics[width=1\textwidth]{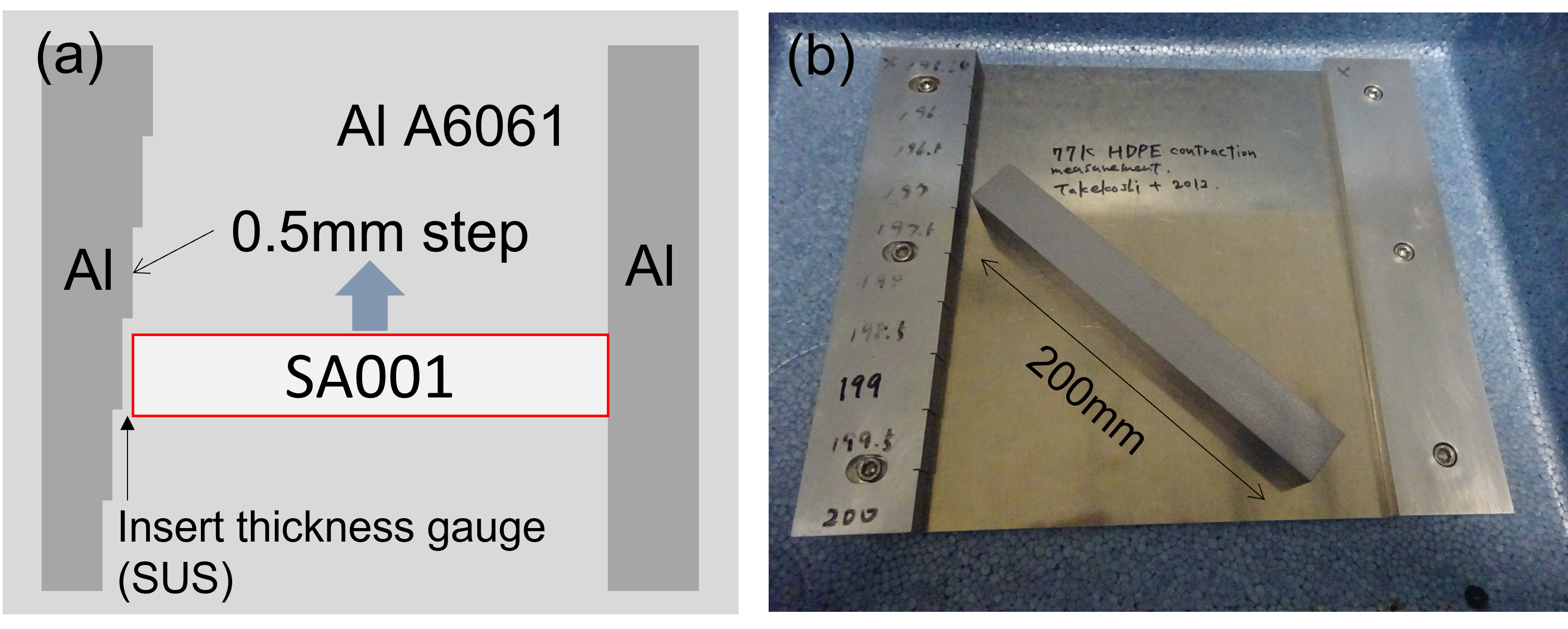}
\caption{(a) Schematic image and (b) photograph of the apparatus and SA001 sample (200$\times 25 \times 25~\mathrm{mm^3}$) for thermal contraction measurement.}
\label{figure:3}
\end{figure}

The thermal contraction of SA001 was measured with a simple apparatus made of 6061 aluminum alloy (Figure~\ref{figure:3}) \citep{2012ITTST...2..584T}.
The distance of the two walls on the aluminum alloy plate, with and without steps, was varied by increments of 0.5~mm.
This functioned as a ruler for measuring the length of the sample.
Both the apparatus and a 200~mm long SA001 sample were immersed in liquid nitrogen.
After fitting the sample to the ruler, a thickness gauge (SUS) was used to accurately determine the sample length.
The contractions of the ruler and the thickness gauge were corrected and the length of the sample was derived.
As a result of the measurement and the contraction correction, we obtained a thermal contraction of  $\frac{L_{293\mathrm{K}}-L_{77\mathrm{K}}}{L_{293\mathrm{K}}}=0.120\pm0.013 \%$, which is equivalent to that of CE7 (0.1\%).

The measured contraction of SA001 at 77~K is can be explained by a volume-weighted mean of contractions (0.116\% was estimated from contractions of silicon and pure aluminum, which are 0.023\%\cite{corruccini1961thermal} and 0.393\% \cite{ekin2006experimental}).
We also estimated the contraction at 4~K to be 0.121\% using the volume-weighted mean of in the same method using the contractions of silicon and aluminum (0.022\%\cite{corruccini1961thermal} and 0.415\%\cite{ekin2006experimental}, respectively).
This is also consistent within the measured contraction at 77~K.

\section{Machining test}
\begin{figure*}
\includegraphics[width=1\textwidth]{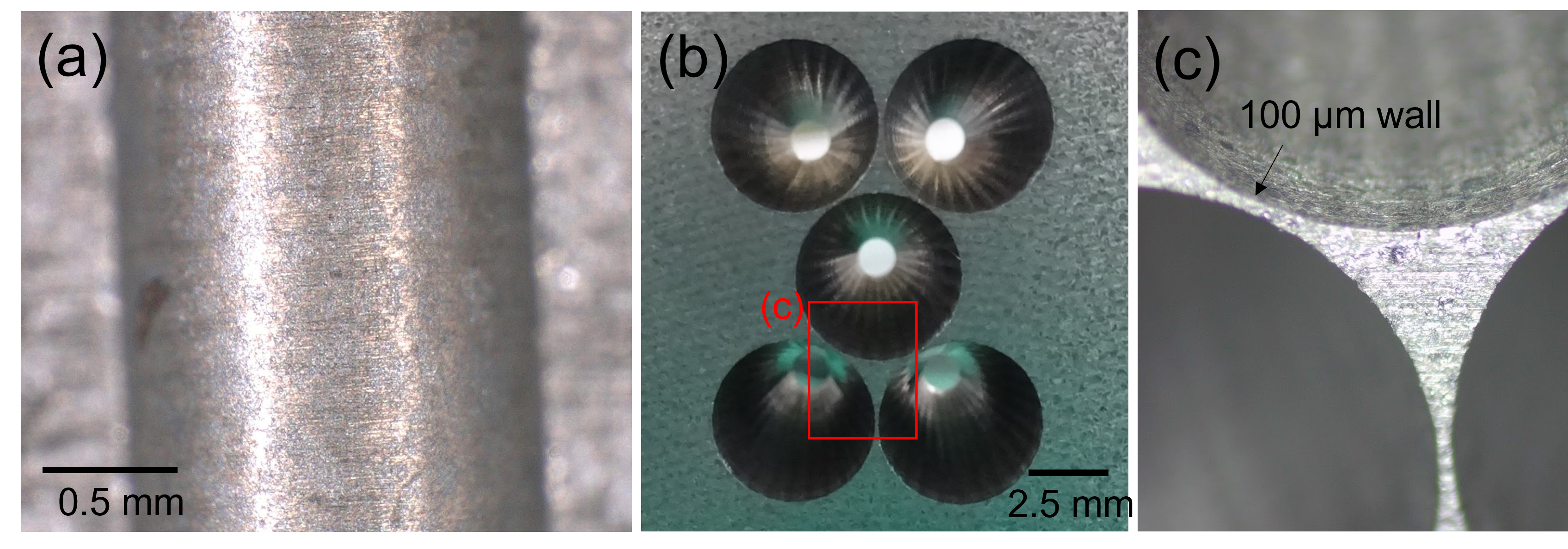}
\caption{(a) Cross section of a circular waveguide. (b) Five-pixel conical feedhorn antennas. (c) Zoom up of the 100~$\mu m$ wall structure between the feedhorns.}
\label{figure:4}
\end{figure*}

Silicon--aluminum alloys are known to be difficult-to-machine materials; therefore, we performed test cuts of circular waveguides and a 5-pixel conical feedhorn on the SA001 samples.
SA001 is a hard and brittle material, and therefore we used diamond-coated tools for machining.
Figure~\ref{figure:4} (a) shows the cross-section of the circular waveguides with a length of 10~mm and a diameter of 1.5~mm.
We confirmed that surface roughness inside the waveguide was achieved to be $R_\mathrm{z} \simeq 1~\mathrm{\mu m}$.

We also fabricated a 5-pixel feedhorn antenna array with an aperture diameter of 5.4~mm, a full flare angle of 15$^\circ$, a waveguide diameter of 1.5~mm, and a total length of 16~mm.
The thin wall structures between the feedhorns were robustly fabricated with a thickness of 100~$\mathrm{\mu m}$.
The machining method is scalable for more than a 100-pixel array.
Thus, we confirmed that SA001 is machinable for conical feedhorn arrays available for large-format superconducting detector instruments.

We also checked the fabrication of more precise structures. Thread holes for M2 and M3 screws can be bored robustly.
Although we tried cutting grooves for a corrugated feedhorn inside the conical horn (groove width $<$0.5~mm), it was still challenging because of the limitation of machining tools and the brittleness of the SiAl composite material.

\section{Conclusion}
\label{sec:conclusion}
We investigated the cryogenic properties of SA001, which are summarized in Table~\ref{table:1}.
The measurement results show that SA001 is suitable for detector packaging or feed horn array material.
We will use SA001 as the material for the focal plane detector modules in a new multi-color millimeter and submillimeter camera planned for the Greenland telescope \cite{2013ASPC..476..243A}.

\begin{acknowledgements}
Measurement samples of SA001 were provided by Japan Fine Ceramics Co., Ltd..
This study was carried out under the Joint Research Program of the Institute of Low Temperature Science, Hokkaido University (21G006, 21G024, 20G013, and 20G033). 
This study was carried out in cooperation with the Advanced Technology Center of the National Astronomical Observatory of Japan (NAOJ).
This work was supported by NAOJ Research Coordination Committee, NINS, Grant Number 2101-0101, and JSPS KAKENHI Grant Number JP17H02872 and JP19K14754.
T.T. is supported by MEXT Leading Initiative for Excellent Young Researchers Grant Number JPMXS0320200188.
S.U. is financially supported by JSPS Research Fellowship for Young Scientists and accompanying Grants-in-Aid for JSPS Fellows (No.21J20742).

\end{acknowledgements}

The datasets generated and analysed during the current study are available from the corresponding author on reasonable request.


\bibliographystyle{spphys}       % APS-like style for physics
\bibliography{reference}   % name your BibTeX data base

%\begin{thebibliography}{}
% etc
%\end{thebibliography}

\end{document}